# Exact Maxwell-Boltzmann, Bose-Einstein and Fermi-Dirac Statistics

Robert K. Niven[1]

[1] School of Aerospace, Civil and Mechanical Engineering, The University of New South Wales at ADFA, Northcott Drive, Canberra, ACT, 2600, Australia.
Email: r.niven@adfa.edu.au

**Abstract**

The exact Maxwell-Boltzmann (MB), Bose-Einstein (BE) and Fermi-Dirac (FD) entropies and probabilistic distributions are derived by the combinatorial method of Boltzmann, without Stirling's approximation. The new entropy measures are explicit functions of the probability and degeneracy of each state, and the total number of entities, $N$. By analysis of the cost of a "binary decision", exact BE and FD statistics are shown to have profound consequences for the behaviour of quantum mechanical systems.

1. **Introduction**

The purpose of this letter is to introduce the exact forms of the degenerate Maxwell-Boltzmann (MB), Bose-Einstein (BE) and Fermi-Dirac (FD) entropy functions, applicable to systems which do not satisfy the Stirling approximation (i.e. in which the total number of entities, $N$, and/or the degeneracy do not approach infinity). They are derived using the combinatorial approach of Boltzmann [1]. The new functions have profound implications. In particular, it is shown that "information" is connected not only with knowledge of the realization (complexion) of a system, but also with knowledge of the number of entities present, and in the case of BE and FD statistics, with knowledge of the degeneracy of each state. The analysis is used to determine the energy cost (in bits) of a "binary decision" for each statistical system, i.e. the cost of learning that a system, distributed over two equiprobable states, is in one such state. The analysis indicates that a binary decision can be purchased for <1 bit in the three systems examined, if one has additional knowledge. Under exact BE statistics, a zero purchase cost is theoretically attainable. However, the cost of a binary decision exceeds 1 bit in BE and FD systems in the absence of such knowledge, except in the Stirling limit. Accordingly, the observation of a BE or FD system is thermodynamically irreversible (requires an energy or information input); it therefore behaves as if it contains infinite entities and is infinitely degenerate until the moment of observation. The analysis provides a rational explanation for the quantum mechanical character of BE and FD systems, and for the "collapse of the wavefunction" in their observation.

The following discussion draws upon the combinatorial definition of entropy [1-5]; the equivalence of information, energy and negative entropy [6-11]; the information-theoretic understanding of the second law of thermodynamics, developed in light of discussions of "Maxwell's demon" [6,8,9,12-15]; and the well-known "double slit experiment" of quantum mechanics [eg, 16].

## 2. Derivations

We start with a variant of Boltzmann's [1] and Planck's [2] combinatorial definition of entropy:

$$\frac{S}{kN} = H = \frac{\ln \mathbb{W}}{N} + C \qquad (1)$$

where $S$ is the total dimensional (thermodynamic) entropy; $k$ is Boltzmann's (or any other applicable) constant; $N$ is the number of entities (e.g., atoms or molecules); $H$ is a dimensionless entropy; $\mathbb{W}$ is the statistical weight or number of possible realizations[1] of the system, of equal probability; and $C$ is an arbitrary constant (reference datum). Note that $H$ is expressed per number of entities present, as is standard practice in information theory [7]. For degenerate MB, BE and FD statistics, the weights have been found to satisfy the following distributions [17; see 3,4]:

$$\mathbb{W}_{MB} = N! \prod_{i=1}^{s} \frac{g_i^{n_i}}{n_i!} = N! \prod_{i=1}^{s} \frac{(\alpha_i N)^{p_i N}}{(p_i N)!} \qquad (2)$$

$$\mathbb{W}_{BE} = \prod_{i=1}^{s} \frac{(g_i + n_i - 1)!}{(g_i - 1)! n_i!} = \prod_{i=1}^{s} \frac{(\alpha_i N + p_i N - 1)!}{(\alpha_i N - 1)!(p_i N)!} \qquad (3)$$

$$\mathbb{W}_{FD} = \prod_{i=1}^{s} \frac{g_i!}{n_i!(g_i - n_i)!} = \prod_{i=1}^{s} \frac{(\alpha_i N)!}{(p_i N)!(\alpha_i N - p_i N)!} \qquad (4)$$

where $i$ denotes each distinguishable state, from a total of $s$ distinguishable states; $n_i$ is the number of entities in each state $i$; $g_i$ is the degeneracy (multiplicity) of each state $i$; $p_i = n_i/N$ is the probability of an entity being in state $i$; and $\alpha_i = g_i/N$ is the *relative degeneracy* of state *i*. Application of Boltzmann's formula (Eq. (1)), assuming that each factorial can be simplified

---

[1] Here the *state* refers to each different category (eg, energy levels or elements) accessible to a system, whilst the *realization* is the physical pattern of the system amongst its states (*complexion* or *microstate*).

using Stirling's approximation:

$$\lim_{x \to \infty}(\ln x!) = x \ln x - x \qquad (5)$$

(whence $N \to \infty$, $n_i \to \infty$, $g_i \to \infty$ and for the FD case $(g_i - n_i) \to \infty$) yields the following degenerate entropy functions [3-5,18]:

$$H_{MB}^{St} \approx -\sum_{i=1}^{s} p_i \ln \frac{p_i}{g_i} \qquad (6)$$

$$H_{BE}^{St} \approx \sum_{i=1}^{s} \left[ (\alpha_i + p_i) \ln(\alpha_i + p_i) - \alpha_i \ln \alpha_i - p_i \ln p_i \right] \qquad (7)$$

$$H_{FD}^{St} \approx \sum_{i=1}^{s} \left[ -(\alpha_i - p_i) \ln(\alpha_i - p_i) + \alpha_i \ln \alpha_i - p_i \ln p_i \right] \qquad (8)$$

where superscript $St$ implies Stirling's approximation. For the BE case (Eq. (7)), the derivation also invokes the limit $N^{-1} \to 0$. The above BE and FD entropies have been expressed previously as functions of $n_i$ and $g_i$ [eg, 3,4], or in terms of the probabilities associated with each degenerate level, $\pi_i = n_i/g_i$ [5,18]. In each case $C=0$ in Eq. (1), and so the above equations do not contain any reference entropy terms.

Due to the differing roles of degeneracy in each statistic, it is most natural to consider the MB entropy as a function of $g_i$, and the BE and FD entropies as functions of $\alpha_i$ (to do otherwise introduces a dependence on infinite $N$). The non-degenerate MB entropy (i.e. $g_i=1$) gives the Shannon entropy, $H = -\sum_{i=1}^{s} p_i \ln p_i$ [7].

Maximization of these degenerate entropy functions by the Lagrangian method in the usual manner, for a physical system subject to natural and energy constraints:

$$\sum_{i=1}^{s} p_i = 1, \quad \sum_{i=1}^{s} p_i \varepsilon_i = \langle \varepsilon \rangle \qquad (9)$$

where $\varepsilon_i$ is the energy of each state $i$ and $\langle \varepsilon \rangle$ is their mathematical expectation, yields the well-known probability distributions [eg, 3-5,18]:

$$p_{MB,i}* \approx g_i e^{-\lambda_0'-\lambda_1\varepsilon_i} = \alpha_i e^{-\lambda_0-\lambda_1\varepsilon_i}, \quad i=1,...,s \tag{10}$$

$$p_{BE,i}* \approx \frac{\alpha_i}{e^{\lambda_0+\lambda_1\varepsilon_i}-1}, \quad i=1,...,s \tag{11}$$

$$p_{FD,i}* \approx \frac{\alpha_i}{e^{\lambda_0+\lambda_1\varepsilon_i}+1}, \quad i=1,...,s \tag{12}$$

where * denotes the most probable realization; $\lambda_0$, $\lambda_1$ are the Lagrangian multipliers associated with the natural and energy constraints respectively, with the same meaning in each statistic, and $\lambda_0' = \lambda_0 + \ln N$. In the MB case, the $\lambda_0$ (or $\lambda_0'$) term can be factored into a partition function $Z = e^{\lambda_0}$ (or $e^{\lambda_0'}$). In all cases $\lambda_0$ and $\lambda_1$ are obtained from the constraints in Eq. (9), with $\lambda_1$ identified with $1/kT$, where $T$ is the absolute temperature [eg, 3,4]. For $e^{\lambda_0+\lambda_1\varepsilon_i} \gg 1$, the BE and FD distributions asymptotically approach the MB distribution.

We now proceed without use of Stirling's approximation. From Boltzmann's definition Eq. (1) and the weights in Eqs. (2)-(4), the exact forms of the MB, BE and FD entropies are:

$$H_{MB}^{x} = \sum_{i=1}^{s}\left\{-\frac{1}{N}\ln[(p_iN)!] + \frac{1}{N}p_i\ln[N!] + p_i\ln g_i\right\} \tag{13}$$

$$H_{BE}^{x} = \sum_{i=1}^{s}\frac{1}{N}\left\{\ln[(\alpha_iN+p_iN-1)!] - \ln[(\alpha_iN-1)!] - \ln[(p_iN)!]\right\} \tag{14}$$

$$H_{FD}^{x} = \sum_{i=1}^{s}\frac{1}{N}\left\{\ln[(\alpha_iN)!] - \ln[(p_iN)!] - \ln[(\alpha_iN-p_iN)!]\right\} \tag{15}$$

In each case we take $C=0$ in Eq. (1); i.e. we are dealing with the "raw" or "exact" forms of each statistic. Note that the external $N!$ in the MB weight (Eq. (2)) is brought inside the summation in Eq. (13). Each entropy function is asymptotic towards its Stirling equivalent (Eqs. (6)-(8)) in the Stirling limits.

Maximization of each entropy subject to the constraints (Eq. (9)) now yields the exact form of each most probable distribution, denoted [#] to distinguish it from its Stirling-approximate

form:

$$p_{MB,i}^{\#} = \frac{1}{N}\left[\Psi^{-1}\left(\frac{\ln[N!]}{N} + \ln g_i - \lambda_0 - \lambda_1 \varepsilon_i\right) - 1\right] \quad (16)$$

$$p_{BE,i}^{\#} = \frac{1}{N}\left[\Psi^{-1}\left[\Psi(\alpha_i N + p_{BE,i}^{\#} N) - \lambda_0 - \lambda_1 \varepsilon_i\right] - 1\right] \quad (17)$$

$$p_{FD,i}^{\#} = \frac{1}{N}\left[\Psi^{-1}\left[\Psi(\alpha_i N - p_{FD,i}^{\#} N + 1) - \lambda_0 - \lambda_1 \varepsilon_i\right] - 1\right] \quad (18)$$

where $\Psi$ is the digamma function and $\Psi^{-1}$ is its inverse, in the latter case invoking the uppermost (positive) branch. Note that the BE and FD distributions are implicit in $p_i^{\#}$.

Although the above distributions are the most probable, owing to the smaller number of realizations they will be less dominant (i.e. the weight $\mathbb{W}$ will be less sharply peaked) than when the Stirling approximation is valid. Similarly, whilst a change in one of the exact entropies (Eqs. (13)-(15)) indicates the average information loss during some process, excursions from this average will be more prominent than in the Stirling-approximate case.

Plots of the partial form (summand) of the non-degenerate ($g_i$=1) exact MB entropy, $H_{MB,i}^{x}$, are shown in Figure 1 for various values of $N$. For brevity, the other entropy functions and probability distributions are not explored here. As shown, $H_{MB,i}^{x}$ follows curves of successively lower entropy with decreasing $N$, each of concave form. The maximum, at probability $p_{MB,i}^{max} = N^{-1}[\Psi^{-1}(N^{-1}\ln[N!]) - 1]$, shifts to higher $p_i$ with decreasing $N$ (for $N \to \infty$, $p_{MB,i}^{max} = e^{-1} = 0.3679$, whereas for $N=1$, $p_{MB,i}^{max} = \Psi^{-1}(0) - 1 = 0.4616$). Note that $H_{MB,i}^{x}$ bears some geometric similarities to the partial forms of the $q$-entropies of Tsallis [19; 20], $H_{Ts} = (q-1)^{-1}\sum_{i=1}^{s} p_i - p_i^q$ for $q>1$, and Abe [21], $H_{Abe} = -(q-1/q)^{-1}\sum_{i=1}^{s} p_i^q - p_i^{1/q}$ for $q \neq 0$ or $\pm 1$, where $q$ is a parameter. However, in contrast to these functions, the exact entropies given here (Eqs. (13)-(15)) are explicitly dependent on $N$.

### 3. Effect of a Binary Decision

Consider now a "binary decision", often examined in information theory, in which we initially have no knowledge of a system containing $N$ entities, except that it consists of two states. We assume the states to be equiprobable, hence $p_i = \frac{1}{2}$, $i = 1,2$. After the decision (i.e. process of observation), the system is found to fully occupy one of the states (say $p_1 = 1$) and not the other ($p_2 = 0$). The information gained by the decision, reflecting our improved state of knowledge, is [6-11]:

$$\Delta I = -(H_{final} - H_{init})/\ln 2 \qquad (19)$$

where $H_{init}$ and $H_{final}$ are the initial and final dimensionless entropies, and the ln 2 divisor converts $\Delta I$ from natural logarithm units ("nats") to binary logarithm units ("bits"). According to the prevailing explanation of "Maxwell's demon" [6,8,9,12-14], in classical (MB) statistics the number of bits of information gained by the decision must be "paid for" by an equivalent number of bits (=$k \ln 2$ units) of energy, i.e.:

$$\Delta E \text{ (bits)} = \Delta I \text{ (bits)} \qquad (20)$$

In this view, the energy-information balance maintains the integrity of the second law of thermodynamics.

In the Stirling limit (i.e. for $N \to \infty$ entities), substitution of parameters for a binary decision into Eqs. (6)-(8), (19) and (20), taking $g_1=g_2=g$ and $\alpha_1=\alpha_2=\alpha$, gives the initial and final entropies and cost in bits, for the MB case:

$$\frac{H_{init,MB}^{St}}{\ln 2} = 1 + \log_2 g \qquad (21)$$

$$\frac{H_{final,MB}^{St}}{\ln 2} = \log_2 g \qquad (22)$$

$$\Delta E_{MB}^{St} = 1 \qquad (23)$$

for the BE case:

$$\frac{H_{init,BE}^{St}}{\ln 2} = \log_2 \frac{(2\alpha+1)^{2\alpha+1}}{(2\alpha)^{2\alpha}} \tag{24}$$

$$\frac{H_{final,BE}^{St}}{\ln 2} = \log_2 \frac{(\alpha+1)^{\alpha+1}}{\alpha^\alpha} \tag{25}$$

$$\Delta E_{BE}^{St} = \log_2 \frac{(2\alpha+1)^{2\alpha+1}}{(\alpha+1)^{\alpha+1}\alpha^\alpha 2^{2\alpha}} \leq 1 \tag{26}$$

and for the FD case:

$$\frac{H_{init,FD}^{St}}{\ln 2} = \log_2 \frac{(2\alpha)^{2\alpha}}{(2\alpha-1)^{2\alpha-1}} \tag{27}$$

$$\frac{H_{final,FD}^{St}}{\ln 2} = \log_2 \frac{\alpha^\alpha}{(\alpha-1)^{\alpha-1}} \tag{28}$$

$$\Delta E_{FD}^{St} = \log_2 \frac{(\alpha-1)^{\alpha-1}\alpha^\alpha 2^{2\alpha}}{(2\alpha-1)^{2\alpha-1}} \geq 1 \tag{29}$$

As evident, a binary decision can be purchased in a MB system for 1 bit (measured per entity), irrespective of degeneracy. In BE and FD systems, the cost also approaches 1 bit as $\alpha \to \infty$. However, at finite $\alpha$, the cost of a binary decision in a BE system is <1 bit, and in a FD system is >1 bit. These curious deviations from expected behaviour are examined shortly.

Now consider a binary decision for which Stirling's approximation does not apply. From Eqs. (13)-(15), (19) and (20), the initial and final entropies and net cost are, for the MB case:

$$\frac{H_{init,MB}^x}{\ln 2} = \frac{1}{N} \log_2 \frac{N!}{\left[(\frac{1}{2}N)!\right]^2} + \log_2 g \tag{30}$$

$$\frac{H_{final,MB}^x}{\ln 2} = \log_2 g \tag{31}$$

$$\Delta E_{MB}^x = \frac{1}{N} \log_2 \frac{N!}{\left[(\frac{1}{2}N)!\right]^2} \tag{32}$$

for the BE case:

$$\frac{H_{init,BE}{}^{x}}{\ln 2} = \frac{2}{N}\log_2 \frac{(\alpha N + \frac{1}{2}N - 1)!}{(\alpha N - 1)!(\frac{1}{2}N)!} \tag{33}$$

$$\frac{H_{final,BE}{}^{x}}{\ln 2} = \frac{1}{N}\log_2 \frac{(\alpha N + N - 1)!}{(\alpha N - 1)!N!} \tag{34}$$

$$\Delta E_{BE}{}^{x} = \frac{1}{N}\log_2 \frac{N!\left[(\alpha N + \frac{1}{2}N - 1)!\right]^2}{\left[(\frac{1}{2}N)!\right]^2 (\alpha N + N - 1)!(\alpha N - 1)!} \tag{35}$$

and for the FD case:

$$\frac{H_{init,FD}{}^{x}}{\ln 2} = \frac{2}{N}\log_2 \frac{(\alpha N)!}{(\alpha N - \frac{1}{2}N)!(\frac{1}{2}N)!} \tag{36}$$

$$\frac{H_{final,FD}{}^{x}}{\ln 2} = \frac{1}{N}\log_2 \frac{(\alpha N)!}{(\alpha N - N)!N!} \tag{37}$$

$$\Delta E_{FD}{}^{x} = \frac{1}{N}\log_2 \frac{N!(\alpha N)!(\alpha N - N)!}{\left[(\frac{1}{2}N)!\right]^2 \left[(\alpha N - \frac{1}{2}N)!\right]^2} \tag{38}$$

For simplicity we take $x! = \Gamma(x+1)$ for non-integer $x$.

For the exact MB case, a plot of the initial and final entropies (Eqs. (30)-(31)) against $N$ for $g=1$ is illustrated in the *information-energy diagram* (in negative entropy units) in Figure 2. As expected (Eq. (21)), for $N \rightarrow \infty$ the binary decision is obtained for a cost of 1 bit. However, as $N$ decreases, the binary decision can be completed for <1 bit, with $\Delta E_{MB}{}^{x}$ diminishing to $\log_2(4/\pi) \approx 0.3485$ bits at $N=1$. This behaviour can be interpreted by considering the information level of $-H_{init,MB}{}^{St}/\ln 2 = -1$ bit, shown as a dotted line in Fig. 2, as a reference datum for the binary decision; at this datum the observer does not know the realization, and also (critically) *does not know N*. In the Stirling limit, lack of knowledge of $N$ has no impact on the binary decision. However, as $N$ decreases, the cost of the binary decision separates into two components: the cost of learning $N$, given by $\Delta E_{MB}{}^{N} = (H_{init,MB}{}^{St} - H_{init,MB}{}^{x})/\ln 2$, and the cost of learning the realization, given by $\Delta E_{MB}{}^{x}$. These costs are labelled in Fig. 2. If one has knowledge of $N \neq \infty$, the binary decision can be completed for a cost of <1 bit; otherwise it is always 1 bit.

Note that for the MB case, the degeneracy has no substantive effect on the binary decision, with both the initial and final entropies being merely shifted by $\log_2 g$ (Eqs. (30)-(31)). The resulting plot is identical to that of Fig. 2, the *y*-axis now representing the information relative to $\log_2 g$. The MB statistic is therefore "well-behaved" in that the total cost of a binary decision, without knowledge of *N*, always sums to 1 bit, irrespective of degeneracy. This is not the case for the other statistics examined.

For the exact BE statistic, the initial and final entropies at α=1000 (which very closely represents α→∞, except for the origin of the *y*-axis scale) are shown in the information-energy diagram in Figure 3a. The reference datum for no knowledge of *N* (here $-H_{init,BE}{}^{St} / \ln 2 \approx -12.409$ bits at α=1000) is also shown. As expected, for *N*→∞ a binary decision can be completed for a cost of 1 bit. However, in contrast to the MB statistic (Fig. 2), both the initial and final entropies curve upwards, such that the cost of learning *N*, given by $\Delta E_{BE}{}^{N} = (H_{init,BE}{}^{St} - H_{init,BE}{}^{x}) / \ln 2$, increases significantly as *N*→1, even as the cost of learning the realization, given by $\Delta E_{BE}{}^{x}$, decreases. In the limit as α→∞ and *N*→1, $\Delta E_{BE}{}^{N} = \log_2(e\pi) - 1 \approx 2.094$ bits, whilst $\Delta E_{BE}{}^{x} = \log_2(4/\pi) \approx 0.3485$ bits. The net effect is that for α→∞ and with knowledge of $N \neq \infty$, a binary decision can be obtained for <1 bit; however, without this knowledge, the total cost is always >1 bit, except in the Stirling limit (*N*→∞).

The effect of degeneracy on the BE statistic is illustrated in Figures 3b-c, respectively for α=1 and α=1/10. In these cases the reference datum represents no knowledge of *N* or the realization, but assumes knowledge of α. From Fig. 3b, if it is known that α=1 and *N*=1, it is possible to complete the binary decision at theoretically zero cost. With knowledge only of α, the total cost is much greater, but becomes <1 bit for $N>15.58$. Similarly, from Fig. 3c, knowing that $\alpha=1/10$ and $N=10$ again yields a theoretically costless process (indeed, a negative cost is predicted for $N<10$, illustrated with dashed lines). With knowledge only of $\alpha$, the total cost of the binary decision is <1 bit for $N>6.52$.

For the FD case, the initial and final entropies at $\alpha=1000$ are virtually identical (as $\alpha\rightarrow\infty$, they are identical) to those for the BE entropy, illustrated in Fig. 3a. Accordingly, with knowledge of $N\neq\infty$, a binary decision can be obtained for <1 bit. Without knowledge of $N$ it is always >1 bit. However, the effect of degeneracy is shown by the plot for $\alpha=1$ in Figure 3d (recall that in FD statistics, $\alpha<1$ has no physical meaning). Evidently, in FD statistics, knowledge of $\alpha$ has the opposite effect as in BE statistics, in that it *increases* the total cost of a binary decision. For $\alpha=1$, without knowledge of $N$, the total cost is exactly 2 bits. With knowledge of $N<2$, it is possible to complete a binary decision for <1 bit; however, the cost is much greater than in the MB and BE cases.

The above analysis indicates that "information" about a physical system is connected not only with knowledge of its realization, but also with knowledge of $N$, and in the case of BE and FD statistics, with knowledge of $\alpha$. With knowledge of $N$ and/or $\alpha$, it is possible to achieve a binary decision at a cost of <1 bit in all systems examined (with some restrictions). Without this knowledge, the cost of a binary decision is always >1 bit in BE and FD systems except in the Stirling limit. BE systems appear to offer the greatest computational advantage, with a zero cost process (and even a negative cost process) being theoretically attainable at low $N$ and $\alpha$. The practical implications of this last finding, for example in the development of quantum computers, warrant further examination.

### 4. Application to Quantum Mechanical Systems

We now turn attention to the famous "double slit experiment" of quantum physics [eg, 16], in which a beam of BE or FD entities (e.g. photons or electrons) is fired through two closely spaced parallel slits towards a detection screen. In bulk, the entities exhibit wave properties, producing an interference pattern on the screen. When only one entity is fired through the slits at a time, it produces a single mark (photographic pixel) on the screen; however, the cumulative pattern of these marks reproduces the interference pattern of a well-populated beam. Evidently, each individual entity appears to behave as if it had passed through both slits, such that the probability of it reaching any point on the screen is the same as when a large number of entities are present. When the entity reaches the screen, its probabilistic distribution collapses to a single observed point, an event referred to as the "collapse of the wavefunction." An attempt to determine the path of each entity - for example using a detector at each slit - causes it to behave as a single (classical) entity; the resulting cumulative probabilistic distribution at the screen is bimodal, with no interference pattern.

In light of the present analysis, it is possible to formulate a new interpretation of the above quantum phenomena (which, like other discussions of these phenomena, is not claimed to be complete). Although an emission device can be set up to emit one BE or FD entity at a time ($N=1$), this knowledge is not "known" to the observer, which in this case is the free space between the emission device and the detection screen. Similarly, the degeneracy of the entity is not known. The reference datum of the observer is therefore akin to the dotted line in Fig. 3a, i.e. the observer has no knowledge of $N$ or $\alpha$. Accordingly, the cost of a binary decision[2] always exceeds 1 bit. In other words, a BE or FD entity carries with it a permanent "budget deficit", such that its observation is thermodynamically irreversible (requires an energy or information input). Since this input is not available, then in accordance with the second law of thermodynamics the entity appears to behave as if $N=\infty$ and $\alpha=\infty$ as it moves from the emission device, through the slits and towards the screen. When it reaches the screen, it triggers a response (e.g. a reaction in a photosensitive cell) which provides the energy or information input needed to "collapse the wavefunction". At this moment, the entity is observed as a single entity. Similar arguments apply to the use of a detector at each slit. The analysis therefore provides a rational explanation for the quantum mechanical character, and the "collapse of the wavefunction", evident in BE and FD systems.

---

[2] Indeed, under exact BE and FD statistics, the cost of a decision between one of $s$ equiprobable alternatives (not analysed here) also exceeds 1 "$s$-bit" (in units of $H^x/\ln s$) except as $N \rightarrow \infty$ or $s \rightarrow \infty$, and so the above analysis is not limited to binary decisions.

The above arguments do not apply to MB systems, which are always "well behaved", requiring only 1 bit of input for a binary decision. Evidently, the distinguishability of MB entities imparts "background information" to the observer, so that their observation (at least in principle) is thermodynamically reversible.

The crux of this analysis is the fact that BE and FD entities are fundamentally indistinguishable. The philosophical basis of quantum behaviour therefore appears to be:

(i) Energy and mass are quantised, taking the form of discrete physical entities;

(ii) Due to Heisenberg's uncertainty principle, it is not possible to simultaneously observe the conjugate variables (e.g. position and momentum) of physical entities, rendering them indistinguishable to an outside observer;

(iii) This indistinguishability gives rise to BE or FD statistics;

(iv) These statistics, applied when $N \not\to \infty$, require that the observation of a physical entity needs the input of energy or information;

(v) In accordance with the second law of thermodynamics, such entities therefore appear to behave as if $N \to \infty$ until their final moment of observation.

## 5. Concluding Remarks

The foregoing analysis brings together information theory, statistical mechanics and quantum theory in a peculiar and very interesting manner. It also reveals a much deeper connection between the combinatorial and information-theoretic bases of the entropy concept than previously believed. Given the importance of the entropy concept, it is surprising that the exact MB, BE and FD entropy functions (Eqs. (13)-(15)) and their implications do not appear to have been examined previously.

The present work has barely scratches the surface of the "exact statistics" derived herein. Several of its features, if correct, will require substantial modifications to previously held beliefs; e.g. the notion that statistical mechanics is meaningful only in the "thermodynamic limit". Also, whilst this study considers the idea of "additional knowledge", how this knowledge is actually held (i.e. the role of the observer) warrants further investigation. Indeed, the analysis hints at the existence of some broader principle of *information relativity*; i.e. systems of different combinatorial character can exhibit different information-theoretic properties, producing an information mismatch between them, with profound consequences for the observation of one system by another. We humans, being necessarily of MB character, are thereby restricted by the second law of thermodynamics to observe the universe through "MB-tinted glasses".

A more detailed discussion of this analysis and its implications will follow elsewhere.

**Acknowledgments**

This work was partly completed during sabbatical leave in 2003 at Clarkson University, New York; McGill University, Quebec; Rice University, Texas and Colorado School of Mines, Colorado, supported by The University of New South Wales and the Australian-American Fulbright Foundation. The author thanks Elanor Huntington, Frank Irons, Colin Pask and the anonymous reviewers for their comments on drafts of the manuscript.

# References

[1] L. Boltzmann, Wien. Ber. 76 (1877) 373, English transl., J. Le Roux (2002), http://www.essi.fr/~leroux/.
[2] M. Planck, Annalen der Physik 4 (1901) 553.
M. Planck, The Theory of Heat Radiation, 2nd ed., Dover, NY, 1959.
[3] R.C. Tolman, The Principles of Statistical Mechanics, Oxford Univ. Press, London, 1938.
[4] N. Davidson, Statistical Mechanics, McGraw-Hill, NY, 1962.
[5] J.N. Kapur, Bull. Math. Ass. India 21 (1989) 39.
[6] L. Szilard, Z. Phys. 53 (1929) 840.
[7] C.E. Shannon, Bell System Technical Journal 27 (1948) 379; 623.
[8] L. Brillouin, Am. Scientist 38 (1950) 594.
[9] L. Brillouin, J. Appl. Phys. 22(3) (1951) 334.
[10] L. Brillouin, J. Appl. Phys. 24(9) (1953) 1152.
[11] M. Tribus, E.C. McIrvine, Scientific Am. 225 (1971) 179.
[12] R. Landauer, IBM J. Res. Dev. 5 (1961) 183.
[13] C.H. Bennett, IBM J. Res. Dev. 17 (1973) 525.
[14] C.H. Bennett, Int. J. Theor. Phys. 21 (1982) 905.
[15] H.S. Leff, A.F. Rex, Maxwell's Demon: Entropy, Information, Computing, Princeton U.P., Princeton, N.J., 1990.
[16] R.P. Feynman, The Feynman Lectures on Physics, Vol. 1, Addison-Wesley, Reading, MA. 1963.
[17] S.N. Bose, Z. Phys. 26 (1924) 178;
A. Einstein, Sitzungsber. Preuss. Akad. Wiss. (1924) 261;
A. Einstein, Sitzungsber. Preuss. Akad. Wiss. (1925) 3;
E. Fermi, Z. Phys. 36 (1926) 902;
P.A.M. Dirac, Proc. Roy. Soc. 112 (1926) 661.
[18] J.N. Kapur, J. Indian Soc. Agric. Stat. 24 (1972) 47;
J.N. Kapur, Indian J. Pure Appl. Math. 14(11) (1983) 1372.
[19] C. Tsallis, J. Stat. Phys. 52(1/2) (1988) 479.
[20] R.K. Niven, Physica A 334(3-4) (2004) 444; see Figure 5a.
[21] S. Abe, Phys. Lett. A 224 (1997) 326; see Figure 1.

**FIGURE CAPTIONS**

Figure 1: Plots of the partial form of the exact degenerate ($g_i=1$) MB entropy $H_{MB,i}^{x}$ against $p_i$ (Eq. (13)), for various values of $N$ (note that as $N \to \infty$, $H_{MB,i}^{x} \to H_{MB,i}^{St}$).

Figure 2: Information-energy diagram for the exact MB statistic, indicating the initial and final entropies and reference datum for a binary decision, for $g=1$.

Figure 3: Information-energy diagrams for (a) exact BE and FD statistics, for $\alpha=1000$ (which closely approximates $\alpha \to \infty$); (b) exact BE statistic, $\alpha=1$; (c) exact BE statistic, $\alpha=1/10$; and (d) exact FD statistic, $\alpha=1$.

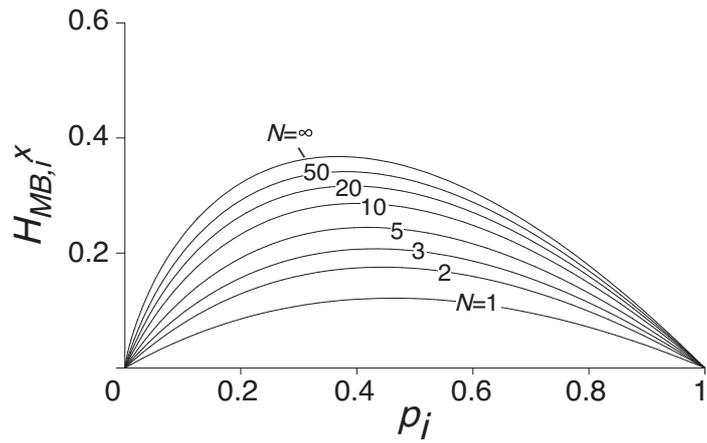

FIGURE 1

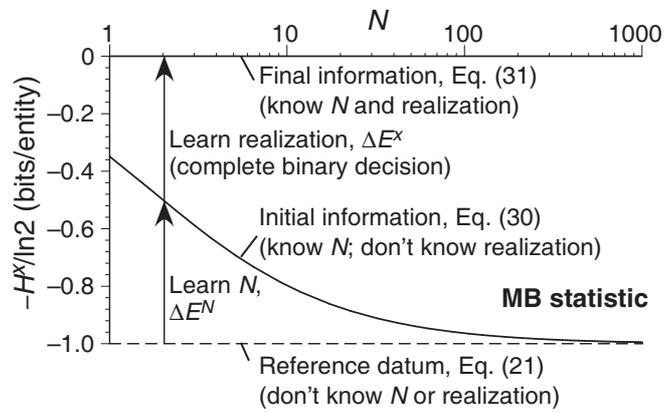

FIGURE 2

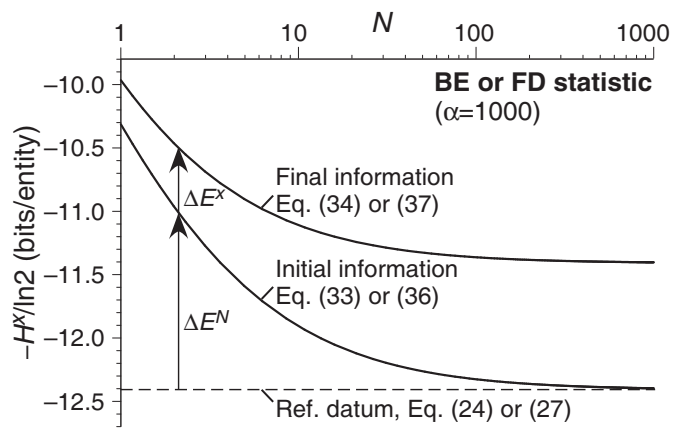

FIGURE 3a

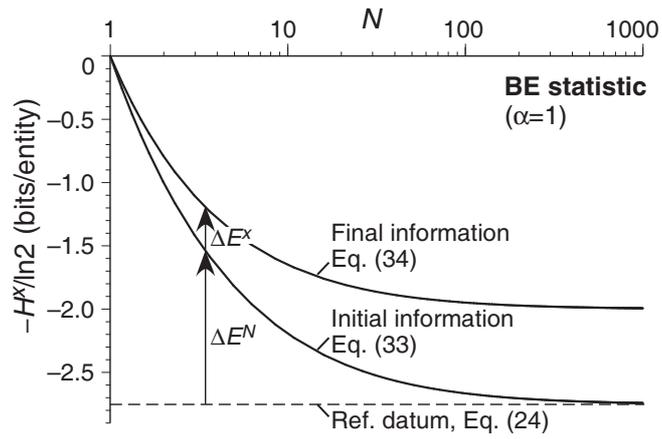

FIGURE 3b

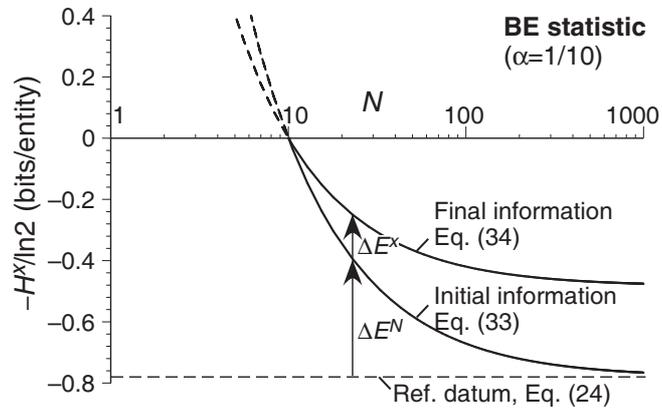

FIGURE 3c

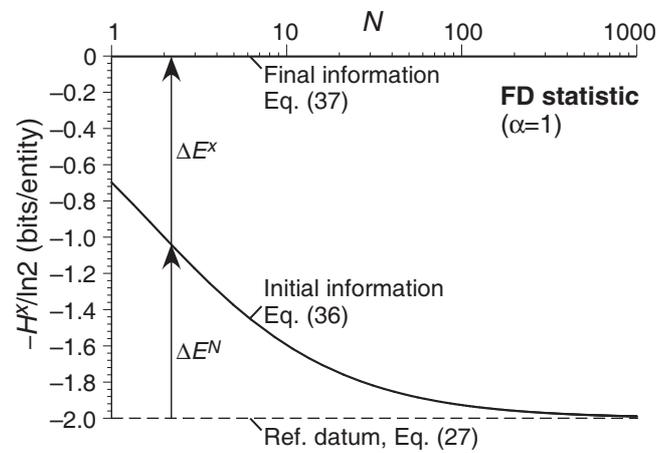

FIGURE 3d

18